\documentclass[prb, superscriptaddress, twocolumn]{revtex4}
\usepackage{graphicx}
\usepackage{enumerate}
\usepackage{hyperref}
\usepackage{textcomp}
\usepackage{amsmath}
\usepackage{amssymb}
\usepackage{pgf}
\usepackage{diagbox}
\usepackage{booktabs}
\usepackage{bbold}
\usepackage{color}
\usepackage{bbm}
\usepackage{soul}

\newcommand{\Tr}{{\mathrm{Tr}}}

\def\ud{\mathrm{d}}

\newcommand{\nep}{\textrm{e}}

\newcommand{\SB}{\mathrm{\scriptscriptstyle SB}}

\newcommand{\OBC}{\mathrm{\scriptscriptstyle OBC}}
\newcommand{\Relax}{\mathrm{\scriptscriptstyle R}}
\newcommand{\Deph}{{\scriptscriptstyle \varphi}}
\newcommand{\Deco}{\mathrm{\scriptscriptstyle D}}

\newcommand{\opbdag}[1]{{\hat{b}^{\dagger}}_{#1}}
\newcommand{\opb}[1]{{\hat{b}^{\phantom \dagger}}_{#1}}
\newcommand{\opcdag}[1]{{\hat{c}^{\dagger}}_{#1}}
\newcommand{\opc}[1]{{\hat{c}^{\phantom \dagger}}_{#1}}
\newcommand{\opeta}[1]{{\hat{\eta}^{\phantom \dagger}}_{#1}}
\newcommand{\opetadag}[1]{{\hat{\eta}^{\dagger}}_{#1}}
\newcommand{\opetatil}{\tilde{\eta}^{\phantom \dagger}}
\newcommand{\opetatildag}{\tilde{\eta}^{\dagger}}

\newcommand{\PauliSigma}{\hat{\sigma}}
\newcommand{\PauliTau}{\hat{\tau}}

\newcommand{\id}{\mathbb{1}}
\newcommand{\ie}{\textit{i.e.} }
\newcommand{\vs}{\textit{vs} }

\newcommand{\kup}{{(k)}}
\newcommand{\sys}{\mathrm{sys}}
\newcommand{\tot}{\mathrm{tot}}
\newcommand{\diss}{\mathrm{diss}}
\newcommand{\coh}{\mathrm{coh}}
\newcommand{\opt}{\mathrm{opt}}
\newcommand{\up}{\mathrm{up}}
\newcommand{\low}{\mathrm{low}}
\newcommand{\defects}{\mathrm{def}}
\newcommand{\thermal}{\mathrm{therm}}
\newcommand{\bath}{\mathrm{bath}}
\newcommand{\full}{\mathrm{full}}
\newcommand{\diag}{\mathrm{diag}}
\newcommand{\Ham}{\widehat{H}}

\newcommand{\enOBC}{\tilde{\epsilon}}

\graphicspath{{Plots/}}

\begin{document}

\title{On the optimal working point in dissipative quantum annealing}

\author{Luca Arceci}
\affiliation{SISSA, Via Bonomea 265, I-34136 Trieste, Italy}

\author{Simone Barbarino}
\affiliation{SISSA, Via Bonomea 265, I-34136 Trieste, Italy}

\author{Davide Rossini}
\affiliation{Dipartimento di Fisica, Universit\`a di Pisa and INFN, Largo Pontecorvo 3, I-56127 Pisa, Italy}

\author{Giuseppe E. Santoro}
\affiliation{SISSA, Via Bonomea 265, I-34136 Trieste, Italy}
\affiliation{CNR-IOM Democritos National Simulation Center, Via Bonomea 265, I-34136 Trieste, Italy}
\affiliation{International Centre for Theoretical Physics (ICTP), P.O. Box 586, I-34014 Trieste, Italy}



\begin{abstract}
We study the effect of a thermal environment on the quantum annealing dynamics of a transverse-field Ising chain. 
The environment is modelled as a single Ohmic bath of quantum harmonic oscillators weakly interacting with the 
total transverse magnetization of the chain in a translationally invariant manner. 
We show that the density of defects generated at the end of the annealing process displays 
a minimum as a function of the annealing time, the so-called {\em optimal working point}, 
only in rather special regions of the bath temperature and coupling strength plane.
We discuss the relevance of our results for current and future experimental implementations with quantum annealing hardware.
\end{abstract}
\maketitle

\section{Introduction} \label{sec:intro}
%
The recent spectacular advancements in the manipulation and control of interacting quantum systems
at the level of a single object, both in equilibrium and far-from-equilibrium conditions, 
opened up a wealth of unprecedented possibilities in the realm of modern quantum physics~\cite{BulutaNori}.
On one hand, they paved the way to the discovery of unconventional states of quantum matter. 
On the other hand, they enabled to exploit quantum mechanics in order to speedup classical computation,
through the implementation of quantum computation or quantum simulation protocols.
In the latter context, one of the most widely known approaches is the so-called
quantum annealing (QA)~\cite{Finnila_CPL94,Kadowaki_PRE98,Brooke_SCI99,Santoro_SCI02},
{\em alias} adiabatic quantum computation~\cite{Farhi_SCI01}.

Due to the realisation of {\it ad-hoc} quantum hardware implementations, mainly based
on superconducting flux qubits, QA is nowadays becoming a field of quite intense
research~\cite{Harris_PRB10,Johnson_Nat11, Denchev_2016, Lanting_2014, Boixo_2013, Dickson_2013, Boixo_2014}.
Its basic strategy works as follows. 
Assume to encode the solution of a given problem in the ground state 
of a suitable Hamiltonian. The goal of the protocol is to find such state by performing
an adiabatic connection (if possible) with another Hamiltonian, typically describing
a much simpler physical system.
Starting from the basic idea rooted on the adiabatic theorem of quantum mechanics~\cite{Messiah:book}, 
a number of different situations that enable a considerable speedup induced by quantum fluctuations
have been extensively analysed in the context of Hamiltonian complexity theory,
in the closed-system setting~\cite{Albash_RMP18}.
Nonetheless, a good description of the physics emerging from the above mentioned experimental devices
cannot neglect the role of dissipation, and the ensuing open-system quantum dynamics~\cite{Weiss:book}.

In the absence of dissipation,
the adiabatic unitary dynamics suggests that a slower annealing will lead to a smaller density of defects 
generated in the process. 
It is a well established fact that, during any non trivial QA dynamics,
one inevitably encounters some kind of phase transition, 
be it a second-order critical point or a first-order transition, where the gap protecting the ground state
--- in principle non-zero, for a finite system --- vanishes as the number of system sites/spins $N$
goes to infinity~\cite{Santoro_JPA06,Zamponi_QA:review, Wauters_PRA17}.
This results in a density of defects $n_{\defects}(\tau)$ that decreases more or less slowly as the annealing time-scale $\tau$ 
is increased. In the second-order case, the predicted decrease of $n_{\defects}(\tau)$ is a power-law $\tau^{-\alpha}$, 
with the exponent $\alpha$ determined by the equilibrium critical point exponents: this is often referred to as
the {\em Kibble-Zurek} (KZ) scaling~\cite{kibble80, zurek96, Polkovnikov_RMP11, Sondhi_2012}. 

Although only marginally considered up to now, still the presence of dissipation modifies this scenario considerably. 
One can argue that an environment will likely have an opposite effect on the density of defects:\cite{Fubini_NJP07} given enough time, 
it would lead to an increase of $n_{\defects}(\tau)$ towards, eventually, a full thermalization in the limit $\tau\to \infty$.
The competing effects of a quantum adiabatic driving in presence of a dissipative environment might therefore lead to interesting 
non-monotonicities in the curve $n_{\defects}(\tau)$: the increase of $n_{\defects}(\tau)$ for larger $\tau$
has been referred to as {\em anti-Kibble-Zurek} (AKZ)~\cite{DelCampo_PRL16}. 
This is, in turn, intrinsically linked to the presence of a {\em minimum} of $n_{\defects}(\tau)$ at some intermediate $\tau_{\opt}$,
known as {\em optimal working point} (OWP). 
It is worth mentioning that the term ``anti-Kibble-Zurek'' appeared for the first time
in a completely classical setting, the adiabatic dynamics of multiferroic hexagonal manganites~\cite{Griffin_PRX12}, 
with the crucial difference 
that the deviation from the expected KZ scenario is there seen as an unexplained decrease of $n_{\defects}(\tau)$ for 
{\em fast} annealings, i.e., for $\tau\to 0$. 
This opposite trend leads to a {\em maximum} of $n_{\defects}(\tau)$ for intermediate $\tau$ 
and, as far as we understand, has nothing to do with AKZ which we will address here in our quantum mechanical framework.

Returning to the quantum case, there have been a few studies on how dissipation affects the QA performance on 
a quantum transverse-field Ising chain in transverse field, where the annealing is performed by slowly switching-off
the transverse field. 
Some studies have employed a classical Markovian noise superimposed to the driving field~\cite{Fubini_NJP07, DelCampo_PRL16, Gao_PRB17} 
or a Lindblad master equation with suitable dissipators~\cite{Gong_SciRep16, Keck_NJP17};
others have considered the effect of one or several bosonic baths coupled to each spin along 
the transverse direction~\cite{Patane_PRL08,Patane_PRB09,Nalbach_PRB15,Smelyanskiy_PRL17}.
%
The general common feature that emerges from these studies is that the density of defects stops following the KZ
scaling after a certain $\tau$, and starts to increase again.

\begin{figure}[!t]
  \centering
  \includegraphics[width=\columnwidth]{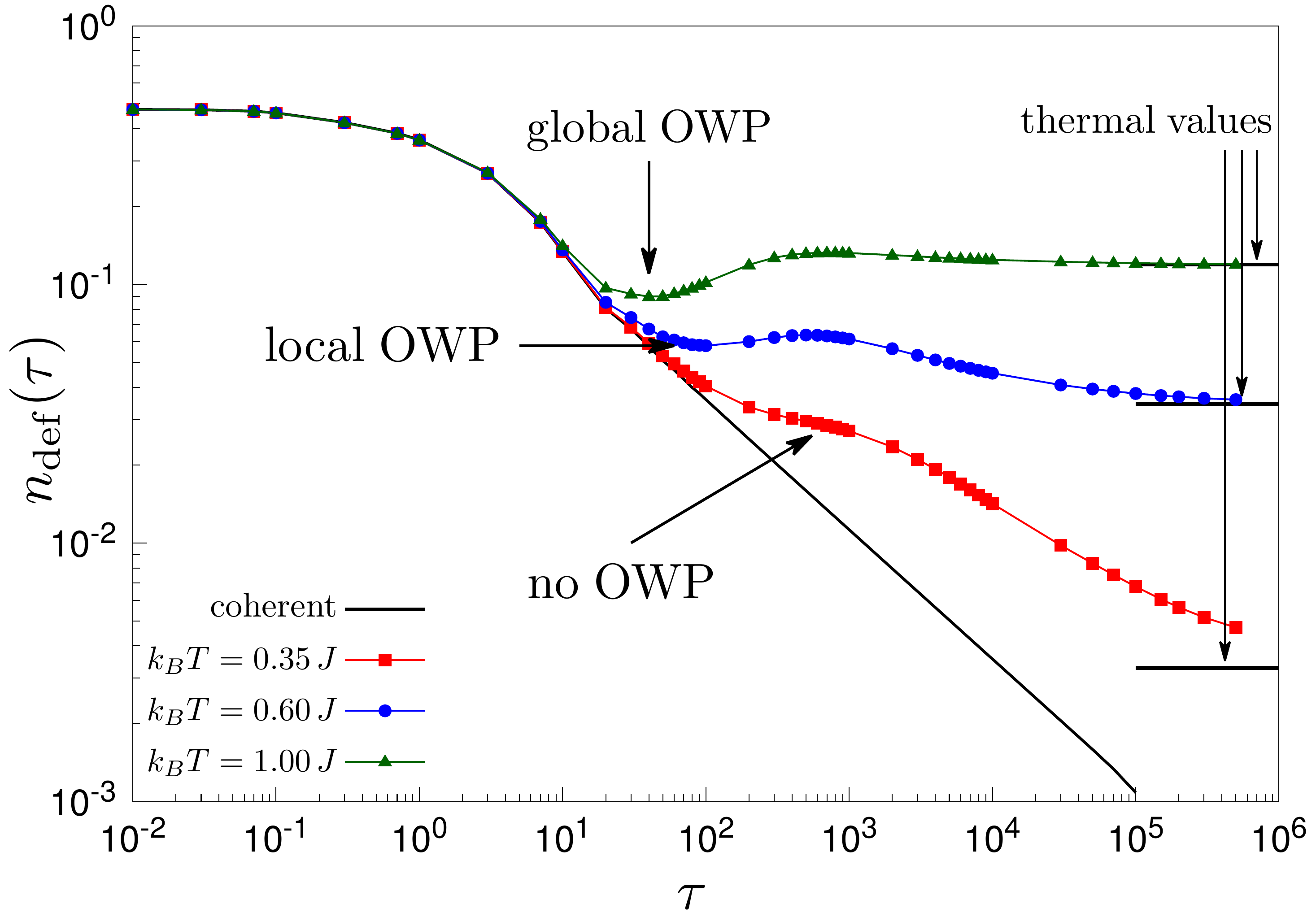}
  \caption{Density of defects \vs annealing time for a quantum Ising chain weakly coupled to an Ohmic bath,
    at different bath temperatures $T$, compared to the ideal coherent evolution (KZ) behaviour, 
    $n_{\defects}(\tau) \sim \tau^{-1/2}$. The plot highlights the three distinct behaviours we have found: 
    $n_{\defects}(\tau)$ can {\it i}) display a global minimum (green triangles), 
    {\it ii}) a local minimum (blue circles), 
    {\it iii}) converge monotonically towards a large-$\tau$ thermal plateau (red squares).
    Here the system-bath coupling constant is kept fixed at $\alpha=10^{-2}$.}
  \label{fig:ndef_tau_intro}
\end{figure}

In this work, we reconsider these issues, concentrating our attention on the benchmark case of a transverse-field Ising chain.
Remarkably, we find that the OWP disappears below a certain temperature, which depends on the system-bath coupling.
The possible situations we encounter are outlined in Fig.~\ref{fig:ndef_tau_intro},
where the density of defects is plotted as a function of the annealing time $\tau$,
for various temperatures and fixed system-bath coupling strengths.
Notice that three different situations may emerge, in which $n_{\defects}(\tau)$ either shows a global
or local minimum at some $\tau_{\opt}$ (\ie the global/local OWP),
and situations where $n_{\defects}(\tau)$ deviates from the simple coherent-dynamics KZ-scaling,
but is still monotonically decreasing, hence no OWP is found.
Quite remarkably, as we shall discuss later on, the range of temperatures that are relevant
for current quantum annealers is such that one would predict the \textit{absence} of an OWP.

We will further comment on the validity of the often used assumption that the density of defects
can be computed as a simple sum of two contributions~\cite{Patane_PRL08, Patane_PRB09, DelCampo_PRL16, Nalbach_PRB15, Keck_NJP17}: 
one given by the purely coherent dynamics, the other coming from a time-evolution governed only by dissipators, 
\ie neglecting the coherent part. 
Since we consider also regimes for which relaxation processes after the critical point are important,
we will provide evidence that this additivity assumption breaks down for large enough annealing times.

The structure of the paper is as follows: 
in Sec. \ref{sec:model} we introduce the dissipative quantum transverse-field Ising chain under investigation,
and discuss the Bloch-Redfield quantum master equation (QME) approach we use to 
work out the dissipative time-evolution of the system.
Our numerical results are illustrated in Sec. \ref{sec:results}: 
we first analyse the conditions for the emergence of an OWP,
also sketching a phase diagram as a function of temperature and system-bath coupling strength.
Next, we investigate the issue of the additivity of the coherent and incoherent contributions
to the density of defects in different regimes.
Finally, in Sec.~\ref{sec:conclusions} we summarize our findings, and provide a discussion of their relevance. 
The two appendices are devoted to a discussion of technical issues related to the QME we have used,
and to the approach towards thermal equilibrium in presence of a bath. 

\section{Model and methods} \label{sec:model}
%
The model we are going to study is described by the following Hamiltonian:
\begin{equation}	\label{H_tot}
	\Ham_{\tot}(t) = \Ham_{\sys}(t) + \Ham_{\SB} + \Ham_{\bath},
\end{equation}
where $\Ham_{\sys}(t)$ is the time-dependent system Hamiltonian, 
$\Ham_{\SB}$ is the system-bath interaction term, and $\Ham_{\bath}$ is a harmonic oscillator 
bath Hamiltonian.
The system is taken to be a quantum spin-$1/2$ Ising chain in a transverse field~\cite{Sachdev:book}:
\begin{equation}	\label{H_XY}
	\Ham_{\sys}(t)  = -J \sum_{i=1}^N \Big[ \PauliSigma_i^x \PauliSigma_{i+1}^x + h(t) \PauliSigma_i^z \Big],
\end{equation}
where $\hat{\boldsymbol{\sigma}}_i \equiv \big( \PauliSigma^x_i, \PauliSigma^y_i, \PauliSigma^z_i \big)$
are the usual Pauli matrices on the $i^\mathrm{th}$ site,
$N$ the number of sites, $J>0$ the ferromagnetic coupling strength,
and $h(t)\geq 0$ the external (driving) field, which is turned off during the evolution.
Periodic boundary conditions (PBC) are assumed, \ie
$\hat{\boldsymbol{\sigma}}_{N+1} = \hat{\boldsymbol{\sigma}}_1$.
The interaction Hamiltonian we considered is written as
%
\begin{subequations}
\begin{align}
	\Ham_{\SB} &= -\frac{1}{2} \hat{X} \otimes \sum_{i=1}^N \PauliSigma_i^z ,\\
	\hat{X} &= \sum_l \lambda_l (\opbdag{l} + \opb{l}) ,
\end{align}
\end{subequations}
where the $\opb{l}$ are bosonic annihilation operators, and $\lambda_l$ are the system-bath coupling constants. 
The bath Hamiltonian is taken, as usual, as $\Ham_{\bath} = \sum_l \hbar \omega_l \, \opbdag{l} \opb{l}$, where
$\omega_l$ are the harmonic oscillator frequencies.
The coupling between the system and the environment is captured by the spectral function~\cite{Leggett_RMP87,Weiss:book} 
$J(\omega) = \sum_l \lambda_l^2 \delta(\omega - \omega_l)$.
We will focus on Ohmic dissipation: for a continuum of frequencies $\omega_l$, we define 
$J(\omega) = 2\alpha \hbar^2 \omega e^{-\omega/\omega_c}$, 
where $\alpha$ quantifies the system-bath coupling strength and $\omega_c$ is a cutoff frequency.
Notice that here we consider a {\em single} (common) bath, which is coupled to all the spins along the $z$-direction,
as done in Ref.~\onlinecite{Nalbach_PRB15}. 
This is, essentially, the quantum version of a noise term acting on the transverse field, whose classical counterpart was treated in Refs.~\onlinecite{Fubini_NJP07, DelCampo_PRL16}. 
This choice of system-bath coupling, which generates infinite-range correlations between all the spins in the chain, 
will allow us to proceed, after further simplifying assumptions, with a simple perturbative QME, as will be clear in a short while.

For the closed system case (\ie without the coupling with the oscillators bath),
the problem can be analytically tackled by means of a standard Jordan-Wigner transformation, 
followed by a Fourier transform~\cite{Lieb_AP61, Pfeuty_AP70},
which allows to rewrite Eq.~\eqref{H_XY} in terms of spinless fermions operators $\opc{k}$ 
in momentum space
\begin{equation} \label{eqn:HsysF}
  \Ham_{\sys}^F = \sum_{k>0} \Big[ \xi_k(t) (\opcdag{k}\opc{k} - \opc{-k}\opcdag{-k} ) +
	\Delta_k (\opcdag{k} \opcdag{-k} + \mathrm{H.c.}) \Big]
\end{equation}
where $\xi_k(t) = 2J \big( h(t) - \cos k \big)$ and $\Delta_k = 2J\sin k$.
The $k$ values in the sum depend on the considered fermionic sector,  
since $\Ham_{\sys}^F$ commutes with the fermion parity~\cite{Dziarmaga_PRL05}. 
The initial ground state belongs to the even-parity sector for any value of the transverse field,
hence the time-evolving state always belongs to that sector.
Due to this fact, one can restrict the choice of the $k$ values to $k=\pi (2n-1)/N$, with $n=1,\ldots, N/2$,
thus fixing the anti-periodic boundary conditions (ABC) for fermions. 

In presence of the system-bath interaction, it is in general not possible to write Eq.~\eqref{H_tot} as a sum of independent 
terms for each given $k$.
However, the fact that the single bath operator $\hat{X}$ couples to a translationally invariant term,
$\sum_i \PauliSigma_i^z = -2\sum_{k>0} (\opcdag{k}\opc{k} - \opc{-k}\opcdag{-k})$, 
ensures momentum conservation for the fermions. 
As argued in Ref.~\onlinecite{Nalbach_PRB15}, this in turn implies that the self-energy for the one-body fermionic 
Green's function is $k$-diagonal, and all fermionic momenta connected to the external lines have momentum $k$. 
Different momenta $k'\neq k$ appear only in closed internal loops.
At the lowest order level (second-order) and within the usual Markovian approximation, the tadpole diagram, which contains a loop, simply 
provides a shift of energy levels and can be neglected, while the only relevant self-energy diagram has momentum $k$ in the fermionic 
internal line \cite{Patane_PRB09}. 
This suggests that, at least at weak coupling and within a Born-Markov approximation, it is legitimate to assume that each momentum 
$k$ does not interact with other momenta $k'\neq k$, and that we could write the coupling to the bath as:
\begin{subequations}
\begin{align}
	\Ham_{\SB} &= \sum_{k>0} \hat{X}_k \otimes (\opcdag{k}\opc{k} - \opc{-k}\opcdag{-k}),  \\
	\hat{X}_k &= \sum_l \lambda_l (\opbdag{l,k} + \opb{l,k}) ,
\end{align}
\end{subequations}
and  $\Ham_{\bath} = \sum_{k>0} \sum_l \hbar \omega_l \, \opbdag{l,k} \opb{l,k}$, where we have effectively 
``split'' the original unique bath into $N/2$ identical copies, one for each fermionic $k$-value, all with identical $J(\omega)$. 
This choice greatly simplifies the problem, since the total Hamiltonian can be written as a sum in $k$-space:
\begin{equation}
  \Ham_{\tot}(t) = \sum_{k>0} \Ham_{k}(t) .
\end{equation}
This automatically leads to an ensemble of independent dissipative two-level systems.
Indeed, it is convenient to map the even-parity fermionic Hilbert space to a collection of pseudo-spin-1/2 quasiparticles, one for each $k>0$,
with the identification $|\!\!\!\uparrow_k\rangle\equiv \opcdag{k}\opcdag{-k}|0\rangle$, and $|\!\!\downarrow_k\rangle\equiv|0\rangle$. 
Introducing the pseudo-spin Pauli matrices
$\hat{\boldsymbol{\tau}}_k \equiv \big( \PauliTau^x_k, \PauliTau^y_k, \PauliTau^z_k \big)$
to represent such two-dimensional space, the Hamiltonian for each $k$ mode reads: 
\begin{equation}
\Ham_{k}(t) = \big( \xi_k(t) + \hat{X}_k \big) \, \PauliTau^z_k + \Delta_k \, \PauliTau^x_k 
+  \sum_l \hbar \omega_l \, \opbdag{l,k} \opb{l,k} .
\end{equation}
Hence, as anticipated, each driven two-level system is coupled with its own bath of harmonic oscillators through a $\PauliTau^z_k$ term.
It is worth to stress that this simplifying assumption of $k$-decoupled baths does not modify the thermal steady state that the system 
reaches at long time, as we will explicitly show in a short while. 

Summarizing the previous discussion, for our specific choice of the system-bath coupling, the dissipative dynamics of 
a translationally invariant quantum Ising chain can be computed by studying the time evolution of $N/2$ two-level systems 
in momentum space, each coupled to an {\em independent identical} bath, described by
a Gibbs density matrix at temperature $T_b = (k_B \beta_b)^{-1}$, where $k_B$ is the Boltzmann's constant:
\begin{equation}
\hat \rho_{\bath} = \frac{\nep^{- \beta_b \Ham_{\bath}}}{\Tr\big\{ \nep^{-\beta_b \Ham_{\bath}} \big\} } \;.
\end{equation}

We address the dissipative dynamics of each two-level system by means of a standard perturbative
Bloch-Redfield QME~\cite{Cohen:book,Gaspard_JCP99a}:
\begin{equation}	\label{Bloch-Redfield}
	\frac{d}{dt} \hat{\rho}_{\sys}^\kup = -\frac{i}{\hbar} \Big[ \Ham_{\sys}^\kup, \hat \rho_{\sys}^\kup \Big] 
	- \Big( \Big[ \PauliTau^z_k, \hat{S}_k(t) \, \hat{\rho}_{\sys}^\kup \Big] + \mathrm{H.c.} \Big) \,,
\end{equation}
where $\Ham_{\sys}^\kup(t) = \xi_k(t) \PauliTau^z_k + \Delta_k \PauliTau^x_k$ is the two-level system Hamiltonian, 
and we assume a weak system-bath coupling and the usual Born-Markov approximation~\cite{Gaspard_JCP99a,Yamaguchi_PRE17}.
Here the first term on the right hand side represents the unitary coherent evolution, while the second term contains the dissipative
effect of the bath on the system dynamics.
The operator $\hat{S}_k(t)$ expresses the interaction of the bosonic bath with the time-evolving
system~\cite{Yamaguchi_PRE17,Arceci_PRB17} in terms of the bath correlation function
$C(t) \equiv C_k(t)=\Tr_{\bath} \Big\{ \hat \rho_{\bath} \, e^{i\Ham_{\bath}t/\hbar} \, \hat{X}_k \,
e^{-i\Ham_{\bath}t/\hbar} \, \hat{X}_k \Big\}$:
%
%
\begin{equation}	\label{S_operator}
\hat{S}_k(t) \approx \frac{1}{\hbar^2} \! \int_{0}^{t}  \ud t' \, C(t') \,
\hat{U}_{0,k}^{\phantom \dagger}(t,t-t') \, \PauliTau^z_k \, \hat{U}_{0,k}^{\dagger}(t,t-t') \;.
\end{equation}
The time-dependence of $\hat{S}_k(t)$ is due to the unperturbed time-evolution operator $\hat{U}_{0,k}(t,t-t')$ 
of the system, which changes with $t$ since $\Ham_{\sys}^\kup(t)$ is driven:
\begin{equation} \label{eq:Ut_evol}
    \hat{U}_{0,k}(t,t-t') = \overrightarrow{\rm Texp} \left[ -\frac{i}{\hbar} \int_{t-t'}^t \ud s \, \Ham_{\sys}^\kup(s) \right] \,.
\end{equation}
Assuming that $C(t)$ decays fast with respect to the time scales of the evolving system,
and that $\Ham_{\sys}^\kup(t)$ is approximately constant on the decay time scale of $C(t)$,
the expression in Eq.~\eqref{eq:Ut_evol} can be drastically simplified.
This allows to write the explicit differential equations that we used to solve for the $\hat{\rho}_{\sys}^\kup(t)$
of each single two-level system, as detailed in Ref.~\onlinecite{Arceci_PRB17}:
we report them in App.~\ref{appA} for the reader's convenience. 

One might wonder how reasonable is our rather special choice of bath in representing the dissipative dynamics of an Ising chain. 
To answer this question, we have looked at the relaxation towards equilibrium at fixed values of the transverse field. 
Any reasonable weakly coupled bath at temperature $T_b$ should allow the system to reach thermal equilibrium values 
for the operators one wants to measure.
This is indeed what the Bloch-Redfield equation~\eqref{Bloch-Redfield} does, but the equilibrium temperature $T$ that 
the system reaches is actually given by $T=T_b/2$, for a reason which is discussed in some detail in App.~\ref{appB}. 
In essence, a peculiarity of our bath-coupling is that {\em only the even-parity fermionic sector} is 
involved in the dissipative dynamics: the odd-parity part of the Hilbert space, which
would correspond to different wave-vectors $k$, is completely decoupled from the bath and neglected altogether. 
This does no justice to the equilibrium thermodynamics, which takes into account {\em all states} in the Hilbert space, and not only 
a dynamically conserved subspace. 
It turns out, however, that accounting for such a part of the Hilbert space simply amounts to having a temperature $T=T_b/2$.
Hence in all the plots, we always indicate the effective temperature $T=T_b/2$ that the system would reach
at thermodynamic equilibrium, rather than the bath-temperature $T_b$ used in our simulations. 

\section{Numerical results} \label{sec:results}
%
Before presenting our results, it is mandatory to introduce the QA protocol we are going to simulate,
and the figure of merit we will use to quantify its performance.
Namely, we choose to vary the external field $h(t)$ in Eq.~\eqref{H_XY}
in the time interval $t \in [0,\tau]$, where $\tau$ denotes the total annealing time,
and implement a standard linear schedule $h(t) = \left(1- t/\tau \right) h_0$, where $h_0$ is the initial value of the field.
In this way, the annealing crosses the zero-temperature critical point of the quantum Ising chain, $h_c=1$,
separating a paramagnetic phase ($h>h_c$) from a ferromagnetically ordered phase ($h<h_c$) in the $\PauliSigma^x$ direction. 
%
%
In all the numerical calculations, we fix the number of sites at $N = 1000$. 
Concerning the bath, we choose $\omega_c = 10J$ as cutoff frequency in the Ohmic spectral function.
The initial condition $\hat{\rho}_{\sys}^\kup(0)$ is chosen to be the ground state of $\Ham_{\sys}^\kup(0)$ for $h(0)=h_0 \gg 1$
(we fix $h_0=10$). 
The time evolution of $\hat{\rho}_{\sys}^\kup(t)$ is then calculated by
integrating the corresponding equations of motion by means of a standard fourth-order Runge-Kutta method.

To assess the quality of the annealing, we compute the average density of defects~\cite{Dziarmaga_PRL05,Caneva_PRB07}
over the ferromagnetic classical Ising state. 
In the original spin language, the operator counting such defects reads:
\begin{equation}	\label{ndef_spin}
	\hat{n}_{\defects} = \frac{1}{2 N} \sum_{i=1}^N \big( 1 - \PauliSigma_i^x \PauliSigma_{i+1}^x \big) \;.
\end{equation}
Translating it into fermions and pseudo-spins, we can write the desired average as:
\begin{equation}	\label{ndef_formula}
	n_{\defects}(t) = \frac{1}{N} \sum_{k>0} \Tr\big\{ \hat{n}_{\defects}^\kup \, \hat{\rho}_{\sys}^\kup(t) \big\}
\end{equation}
where $\hat{n}_{\defects}^\kup = \id - \PauliTau^z_k \cos k + \PauliTau^x_k \sin k$. 
%

In the following, we discuss the dependence of the final density of defects $n_{\defects}(t=\tau)$ on the annealing time $\tau$, 
for different system-bath coupling strengths $\alpha$ and temperatures $T=T_b/2$. 
In particular, we characterise the regimes for which an OWP is present or not,  
and study how the defect density approaches thermal values for long annealing times.
We also analyse the conditions under which the processes of coherent and incoherent defect production 
can be regarded as independent, and highlight regimes in which this assumption fails.

\subsection{The optimal working point issue}
%
\begin{figure}[tbp]
  \centering
  \includegraphics[width=\columnwidth]{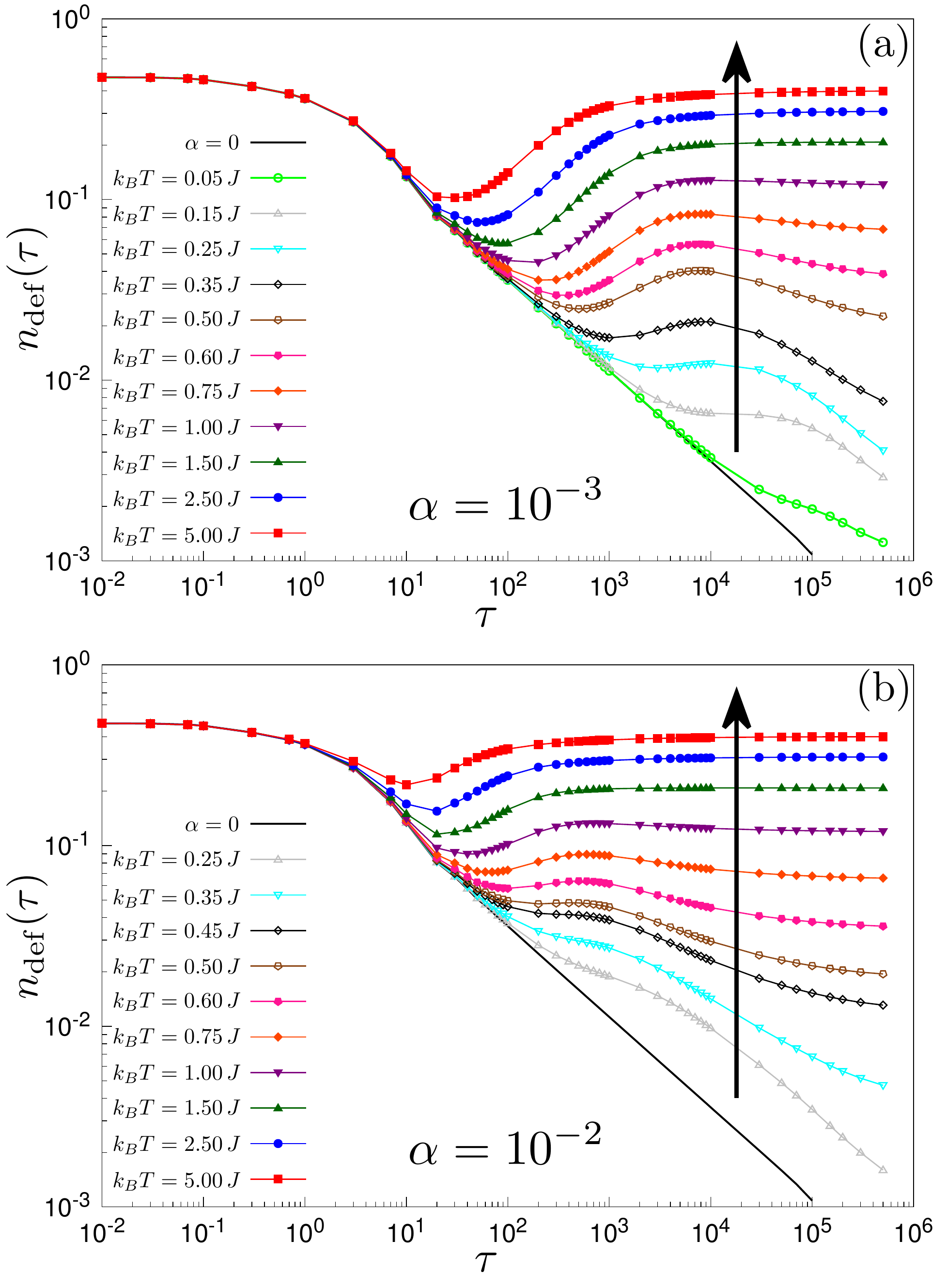}
  \caption{Density of defects \vs annealing time $\tau$ for (a) $\alpha = 10^{-3}$, (b) $\alpha = 10^{-2},$ 
  for different effective temperatures $T$, as indicated in the legend. 
  The arrows indicate the direction of increasing temperatures.
  The trend for high $T$ is of AKZ type, with an emergent OWP. 
  At lower $T$ and/or higher $\alpha$ values a monotonic trend smoothly appears, with the absence of OWP.}
  \label{fig:ndef_tau}
\end{figure}
%
\begin{figure}[tbp]
  \centering
  \includegraphics[width=\columnwidth]{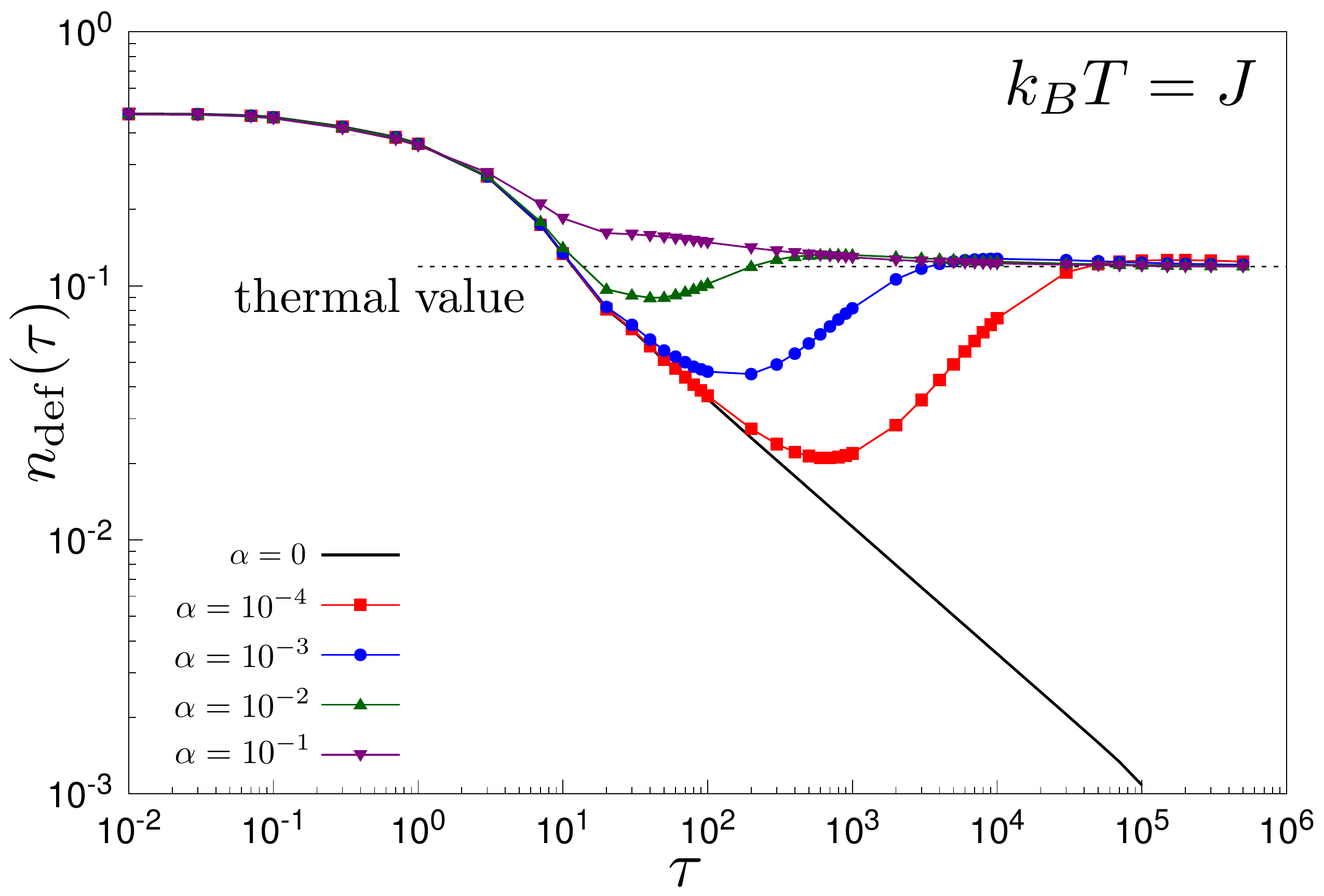}
  \caption{Density of defects \vs annealing time $\tau$ for $k_BT = J$, at different coupling strengths $\alpha$. 
   Note that, for large enough annealing times, $n_{\defects}(\tau)$ converges towards the thermal value.}
  \label{fig:ndef_tau_T}
\end{figure}
%
%
\begin{figure}[tbp]
  \centering
  \includegraphics[width=\columnwidth]{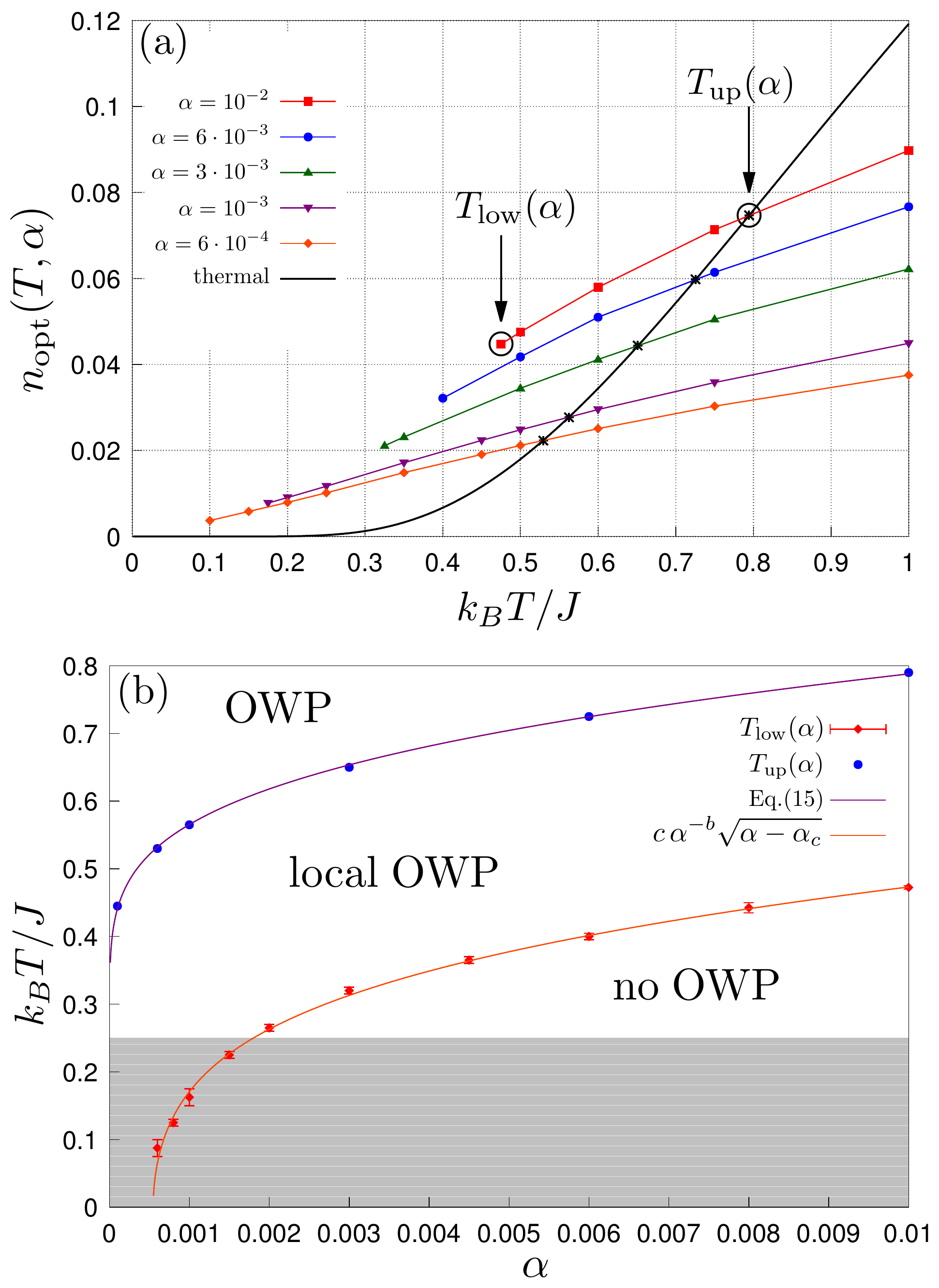}
 \caption{(a) Dependence of $n_{\opt}$ on $T$, for various values of $\alpha$. 
 Each curve defines an upper value $T_{\up}(\alpha)$ at which $n_{\opt}(T_{\up})=n_{\infty}(T_{\up})$
 (marked by stars), and a lower $T_{\low}(\alpha)$ at which the local minimum defining $n_{\opt}$ disappears.
 \newline
 (b) Phase diagram in the $T-\alpha$ plane with $T_{\up}(\alpha)$ and $T_{\low}(\alpha)$. 
 A proper OWP only exists for $T> T_{\up}(\alpha)$. 
 The shaded area alludes to the typical range of temperatures of interest for the 
 D-Wave$^{\mbox{\tiny \textregistered}}$ hardware~\cite{Harris_PRB10,Johnson_Nat11},  
 with $k_BT\simeq 12$ mK and $J\gtrsim 80$ mK. 
  }
  \label{fig:phasediag}
\end{figure}
%
%
Let us start by looking at the behaviour of the final density of defects $n_{\defects}(\tau)$ as a function of the annealing time $\tau$.  
In Fig.~\ref{fig:ndef_tau} we consider $\alpha = 10^{-3}$ and $10^{-2}$, for which our perturbative approach 
is reliable~\cite{Arceci_PRB17}, and different values of $T$.
For sufficiently high temperatures, we observe a clear AKZ trend: after the initial decrease,
$n_{\defects}(\tau)$ attains an absolute minimum at some
value $n_{\opt}=n_{\defects}(\tau_{\opt})$ --- corresponding to the OWP $\tau_{\opt}$ --- 
and then starts to increase again towards a large-$\tau$ plateau at $n_{\infty}=n_{\defects}(\tau\to\infty)$.
By decreasing $T$, however, the plateau value $n_{\infty}$ can become smaller than $n_{\opt}$, hence $\tau_{\opt}$ would
correspond to a {\em local minimum} and should be called, strictly speaking, a ``{\em local} optimal working point''.  
A further reduction of $T$ leads to the disappearance of the local minimum at $\tau_{\opt}$, with a monotonic decrease of 
$n_{\defects}(\tau)$ as $\tau$ grows.
By comparing the two plots, it is clear that the presence of an OWP is determined by the interplay between 
the temperature $T$ and system-bath coupling strength $\alpha$.
%

Conversely, Fig.~\ref{fig:ndef_tau_T} displays the final density of defects
for a fixed temperature, $k_BT=J$, while scanning $\alpha$ in the range $[10^{-4}, 10^{-1}]$: 
we see that very weak couplings favour an AKZ behaviour, while stronger couplings tend to
cause the lack of an OWP. 
Moreover, it appears neatly that $n_{\defects}(\tau)$ exhibits a convergence, for large $\tau$,
towards a value $n_{\infty}(T)$ which depends {\em only} on the temperature $T$.
We have verified that such limiting value coincides with the final ($h=0$) thermal value $n_{\thermal}(T)\equiv n_{\defects}^{T}(h=0)$,
indicated by a horizontal dashed line in Fig.~\ref{fig:ndef_tau_T} and calculated from the equilibrium average:
\begin{equation} \label{eqn:thermal_t}
n_{\defects}^{T}(h) = \frac{1}{N} \sum_{k>0} \Tr\big\{ \hat{n}_{\defects}^\kup \, \hat{\rho}_{\sys}^{T}(h) \big\} \;,
\end{equation}
where $\hat{\rho}_{\sys}^{T}(h)$ is the system thermal state at bath temperature $T_b=2T$, when the transverse field is $h$. 
The explicit calculation of $n_{\thermal}(T)$, following App.~\ref{appB}, brings:
\begin{equation}
n_{\infty}(T) \equiv n_{\thermal}(T) =  \frac{1}{2} \big( 1-\tanh{(\beta J)} \big) \;.
\end{equation}
Figure~\ref{fig:phasediag}(a) summarizes the values obtained for $n_{\opt}(T)$ versus $T$, for various $\alpha$.
The stars mark the temperatures $T_{\up}(\alpha)$ where $n_{\opt}(T)$ crosses the  (infinite-time limit) thermal value $n_{\thermal}(T)$:
given $\alpha$, only for $T> T_{\up}$ the minimum at $\tau_{\opt}$ is an absolute minimum of $n_{\defects}(\tau)$.
For $T<T_{\low}(\alpha)$ the minimum disappears completely --- $n_{\defects}(\tau)$ is a monotonically decreasing function of $\tau$. 
For $T_{\low}(\alpha)<T<T_{\up}(\alpha)$, $n_{\opt}$ survives only as a local minimum.
Summarizing, for the range of $\alpha$ we have investigated (the weak-coupling region $\alpha<10^{-1}$)
one can construct two characteristic temperature curves, $T_{\low}(\alpha)<T_{\up}(\alpha)$ and a phase diagram, sketched
in Fig.~\ref{fig:phasediag}(b). 
Notice that the two curves are difficult to extrapolate from the data for $\alpha\to 0$, because the simulations
would require a too large time-scale to observe the presence or absence of the local minimum in $n_{\opt}$. 
We can however argue, on rather simple grounds, that $T_{\up}(\alpha\to 0)$ should drop to zero as $\sim 1/\log(1/\alpha)$. 
Indeed, as seen from Fig.~\ref{fig:phasediag}(a), $n_{\opt}(T,\alpha)$ appears to be roughly linear in $T$ in the region 
where it crosses the thermal curve, $n_{\opt}(T,\alpha) \simeq A_{\alpha} T$, with a slope $A_{\alpha}$ which, as we have
verified, depends on $\alpha$ in a power-law fashion.
Since $n_{\thermal}(T)\sim \nep^{-2J/k_BT}$ for small $T$, we can write the implicit relationship:
\begin{equation} 
n_{\opt}(T_{\up},\alpha) \simeq A_{\alpha} T_{\up} \simeq  \nep^{-2J/k_BT_{\up}} \;. 
\end{equation}
Assuming a power-law for $A_{\alpha}$ we get, up to sub-leading corrections,
\begin{equation}
T_{\up}(\alpha) \sim \frac{\mathrm{const}}{\log{\frac{1}{\alpha}} + O\left( \log \log \frac{1}{\alpha} \right)} \;.
\end{equation}
This functional form fits our numerically determined data in a remarkably good way. 
The behaviour of the $T_{\low}(\alpha)$ curve is considerably less trivial. 
On the practical side, it is computationally harder to obtain information of the temperature below which a local OWP ceases to exists.  
Our data suggest that there might be a critical value $\alpha_c\approx 5.5 \cdot 10^{-4}$ below which a local OWP exists 
even at the smallest temperatures, but this might be an artefact of some of the approximations involved in our weak-coupling QME. 
All in all, the phase diagram is quite clear --- at least for weak-moderate values of $\alpha$ --- in predicting the presence of a
true OWP only for relatively large temperatures $T$. We will discuss this in the concluding section. 

\begin{figure}
  \centering
  \includegraphics[width=\columnwidth]{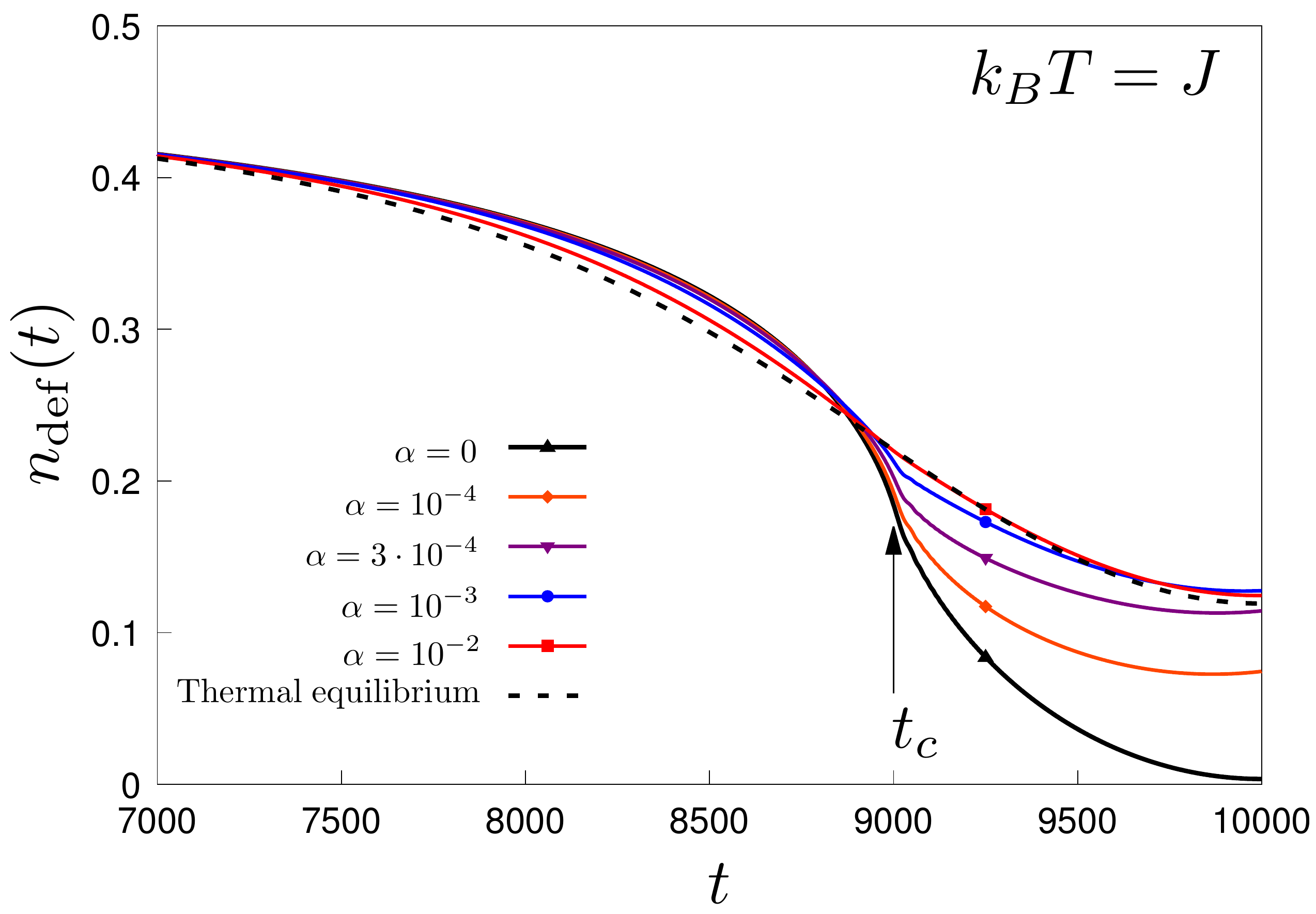}
  \caption{Density of defects \vs time for $\tau = 10^5$, $k_BT = J$ and different system-bath coupling strengths.
  The arrow at $t_c$ marks the value of $t$ at which the transverse field crosses the critical value, $h(t_c)=h_c$.  
  For $\alpha = 10^{-2}$, where the defects density has fully converged (see Fig.~\ref{fig:ndef_tau_T}), $n_{\defects}(t)$
  is almost superimposed to the exact instantaneous thermal one computed from Eq. \ref{ndef_thermal}.
  }
  \label{fig:ndeftime_T2}
\end{figure}
The fact that the system converges to a thermal state for long annealing times is quite reasonable, and perhaps
expected.
Indeed, if the thermalization time-scale becomes smaller than the annealing time-scale, one would expect that the system's state remains 
close to the instantaneous thermal equilibrium state at every time during the whole dynamics.
Figure~\ref{fig:ndeftime_T2}, where we plot $n_{\defects}(t)$ \vs time at fixed $k_BT = J$ and fixed annealing time $\tau = 10^5$, confirms
this expectation.  

%
In Fig.~\ref{fig:ndeftime_T2} 
the dashed line indicates, as a guide, the ``instantaneous'' exact thermal value $n_{\defects}^{T}(h(t))$
computed according to Eq.~\eqref{eqn:thermal_t} (see App.~\ref{appB} for details),
while the arrow at $t_c$ marks the value of $t$ where the transverse field $h(t)$ crosses the critical point, $h(t_c)=h_c=1$. 
We observe that, 
for increasing couplings $\alpha$, the curves tend to be closer and closer to the instantaneous thermal one, 
since the thermalization time-scale decreases. 
%

\subsection{Interplay between coherent and incoherent defects production}
As mentioned above, in absence of dissipation, the defects produced are due to violations of adiabaticity in the coherent dynamics
\begin{equation} \label{Bloch-Redfield-coh}
\frac{d}{dt} \hat{\rho}_{\coh}^\kup(t) = -\frac{i}{\hbar} \Big[ \Ham_{\sys}^\kup(t), \hat \rho_{\coh}^\kup(t) \Big]  \;,
\end{equation}
and would be given by:
\begin{equation}	
n_{\defects}^{\coh}(t) = \frac{1}{N} \sum_{k>0} \Tr\big\{ \hat{n}_{\defects}^\kup \, \hat{\rho}_{\coh}^{\kup}(t) \big\} \;.
\end{equation}
As well known, $n_{\defects}^{\coh}(t=\tau)$ obeys the usual KZ scaling~\cite{Polkovnikov_RMP11}.
In the present case, for the Ising chain, $n_{\defects}^{\coh}(\tau)\sim \tau^{-1/2}$.

In the literature related to dissipative QA, it is often found that the density of defects can be 
regarded as the sum of two independent contributions
\begin{equation} \label{eqn:additivity}
	n_{\defects}(t) \approx n_{\defects}^{\coh}(t) + n_{\defects}^{\diss}(t) \;.
\end{equation}
%
The second contribution, $n_{\defects}^{\diss}(t)$, should be due to a {\em purely dissipative} time-evolution of the system state:
\begin{subequations}	\label{Bloch-Redfield-diss}
\begin{align}
  \frac{d}{dt} \hat{\rho}_{\diss}^\kup &= 
  - \Big( \big[ \PauliTau^z_k, \hat{S}_k(t) \, \hat{\rho}_{\diss}^\kup \big] + \mathrm{H.c.} \Big), \\
  n_{\defects}^{\diss}(t) &= \frac{1}{N} \sum_{k>0} \Tr\big\{ \hat{n}_{\defects}^\kup \, \hat{\rho}_{\diss}^{\kup}(t) \big\} \,.	
\end{align}
\end{subequations}
Notice that the time evolution of the system Hamiltonian enters here only through the bath-convoluted system operator
$\hat{S}_k(t)$. 
In particular, based on this ``additivity'' assumption, Refs.~\onlinecite{Patane_PRL08, Patane_PRB09, Nalbach_PRB15} have 
computed scaling laws for the defects density in presence of dissipation due to one or more thermal bosonic baths.
However, a crucial requirement for these scaling laws to hold is that  thermalization effects after the critical point has been crossed 
must be negligible: indeed, in Refs.~\onlinecite{Patane_PRL08, Patane_PRB09}, the adiabatic sweep is stopped {\em at the critical point} 
or immediately below it, so giving no time to the system to ``feel'' the thermal environment; 
in Ref.~\onlinecite{Nalbach_PRB15}, the analysis is carried out for a very small system-bath coupling $\alpha$, so that the 
thermalization time is extremely long, much longer than the characteristic annealing time scale. 
As a consequence, after the critical point crossing, the system is very weakly affected by the bath, and the additivity 
assumption still holds.

Here we are considering an annealing protocol that can leave enough time to the system to thermalize after the critical point crossing: 
indeed, the quantum critical point is crossed when $h(t_c)=(1-t_c/\tau)h_0=1$, in our units, hence
$t_c = (1-1/h_0) \tau= 0.9\tau$, for $h_0=10$. 
This means that, after the critical point, the system has $t_{\mathrm{avail}} = 0.1\tau$ time to relax to the thermal state, 
\ie a time proportional to the annealing time $\tau$. 
Therefore, for all the $\tau$ values for which $t_{\mathrm{avail}}$ is comparable or larger than the bath thermalization time, 
the effect of the bath after the quantum critical point will be no more negligible.

\begin{figure}
  \centering
  \includegraphics[width=\columnwidth]{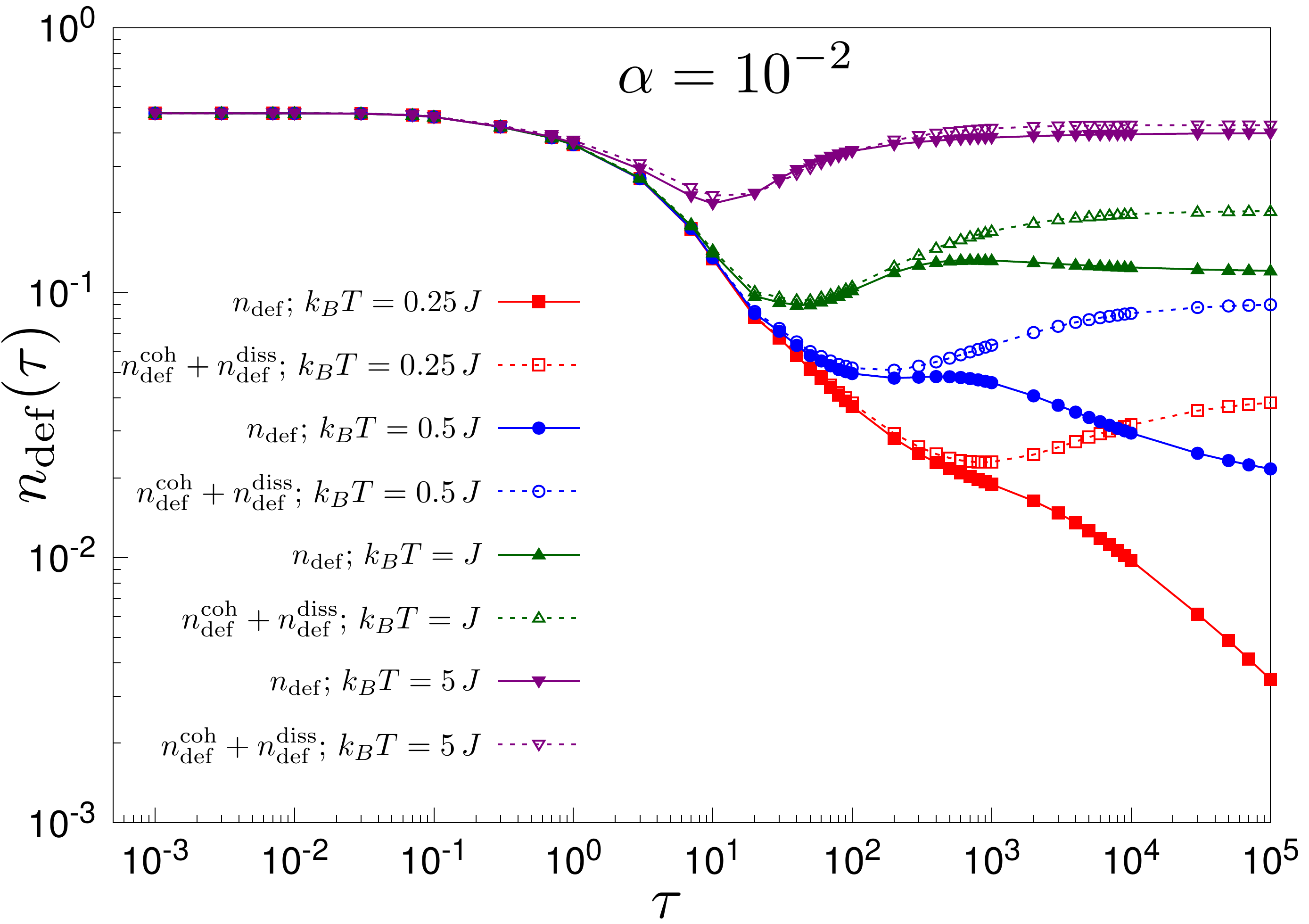}
  \caption{Test for the additivity assumption Eq.~\eqref{eqn:additivity} for the formation of defects.
    We compare $n_{\defects}(t=\tau)$, calculated with the full Bloch-Redfield evolution
    in Eq.~\eqref{Bloch-Redfield} (continuous curves, filled symbols), to the sum of $n_{\defects}^{\coh}(\tau)$
    plus the purely-dissipative evolution contribution $n_{\defects}^{\diss}(\tau)$ from Eq.~\eqref{Bloch-Redfield-diss}
    (dashed curve, empty symbols).}
  \label{fig:additivity}
\end{figure}

Figure~\ref{fig:additivity} shows a test of the additivity assumption for four different bath temperatures at fixed coupling $\alpha = 10^{-2}$; 
for each temperature, we compare the defects density obtained via Eq.~\eqref{Bloch-Redfield} with that obtained by the sum of $n_{\defects}^{\coh}$ and $n_{\defects}^{\diss}$.
For $\tau$ small enough, the additivity assumption always holds, since $t_{\mathrm{avail}}$ is too short, \ie there is not enough time 
to feel the effect of the bath after the critical point is crossed.
However, for longer annealing times the additivity starts to fail: the lower the temperature, the worse it is.
In particular, we see that additivity would {\em always} predict the presence of an OWP, but in some regimes the interplay between coherent and dissipative effects is non-trivial and the two contributions cannot be considered separately.
Note also that for $k_BT = 5J$ 
the additivity assumption seems to hold for every annealing time, even after converging 
to its thermal value. However, this is probably due to the fact that both two values tend to converge to the maximum
for the density of defects, and therefore additivity holds better.

\section{Discussion and conclusions} \label{sec:conclusions}
%
In the present paper, we have revisited some of the issues related to QA in presence of dissipation. 
In particular, we investigated under which conditions it is possible to find an ``optimal'' annealing time,
the optimal working point (OWP), that minimizes the number of defects, and therefore maximizes the annealing performance.
%
We have tackled those issues in the benchmark case of a transverse-field Ising chain QA, by studying its open-system
quantum dynamics with a Markovian QME, as appropriate for a dissipative environment modelled by a standard
Caldeira-Leggett Ohmic bath, weakly coupled in a uniform way to the transverse magnetization.
Of course, such a choice of the system-bath coupling is rather peculiar and very specific.
However, we have tested that it provides the correct steady-state thermalization for a chain evolving at fixed 
transverse field; hence we expect that our results should retain some general validity, at least qualitatively,
for other forms of thermal bosonic baths. 

Interestingly, a proper OWP can be seen essentially only in a high-temperature regime, $k_BT\gtrsim 0.5 J$. 
For temperatures which might be relevant for current~\cite{Harris_PRB10,Johnson_Nat11}, and presumably future
quantum annealers, $k_BT \ll J$, schematically sketched by a shaded area in the phase-diagram of Fig.~\ref{fig:phasediag}(b), 
we found that $n_{\defects}(\tau)$ would be monotonically decreasing (hence without OWP), 
except for very weak bath couplings, $\alpha\lesssim 10^{-3}$. 
In the intermediate temperature regime, $n_{\defects}(\tau)$ displays a {\em local minimum} at finite $\tau$, 
but the actual global minimum is attained as a $\tau\to\infty$ thermal plateau.
%
Obviously, the previous considerations would apply to experimental realizations where the coupling to the environment can be
considered to be weak and Ohmic, which apparently is not the case for the D-Wave$^{\mbox{\tiny \textregistered}}$
hardware~\cite{Harris_PRB10,Johnson_Nat11}, where $1/f$ noise seems to play an important role~\cite{Boixo_2016}. 
The extension of our study to cases where the bath spectral density has different low-frequency behaviours, 
such as sub-Ohmic or with $1/f$ components, is a very interesting open issue which we leave to a future work. 

Previously related studies~\cite{Patane_PRL08, Patane_PRB09, Nalbach_PRB15} on the same model
did not detect all these different behaviours, because they either stopped the annealing close to the critical
point~\cite{Patane_PRL08, Patane_PRB09} --- to highlight some universal aspects of the story, which survive in presence of the environment --- 
or considered an extremely small, $\alpha\sim 10^{-6}$, system-bath coupling~\cite{Nalbach_PRB15}:
this amounts, in some sense, to effectively disregarding thermalization/relaxation processes occurring after the critical point 
has been crossed. 

A second issue we have considered is the additivity {\em Ansatz} on the density of defects, Eq.~\eqref{eqn:additivity}, 
i.e., its being a simple sum of the density of defects coming from the coherent dynamics with that originating from the time evolution 
due to dissipators only:
we found that additivity breaks down as soon as the bath thermalization time is effectively shorter than the characteristic 
time-scale for the system dynamics; for our annealing protocol, this happens at long enough annealing times 
$\tau$, as shown in Fig.~\ref{fig:additivity}.

In conclusion, we believe that QA protocols realized with quantum annealers for which thermal effects are sufficiently weak, 
at sufficiently low temperatures, should not show any OWP, but rather a monotonic decrease of the error towards a large running
time thermal plateau.
%
Furthermore, it would be tempting to move away from the reference quantum Ising chain toy-model,
and explore the effects of dissipation in more sophisticated models.
The use of quantum trajectories~\cite{Daley_AP14} or of tensor-network approaches,
  recently extended to deal with open quantum systems~\cite{Verstraete_04, Zwolak_04}, could help in addressing generic
  one-dimensional (or quasi-one-dimensional) systems, which would be hardly attackable from an analytic perspective.

\section*{ACKNOWLEDGMENTS}
We acknowledge fruitful discussions with A. Silva, R. Fazio and G. Falci.
Research was partly supported by EU FP7 under ERC-MODPHYSFRICT, Grant Agreement No. 320796.

\vspace{3mm}

\appendix
\section{Differential equations from the Bloch-Redfield approach} \label{appA}
%
Following the procedure used in Ref.~\onlinecite{Arceci_PRB17}, we start from Eq.~\eqref{Bloch-Redfield} and apply a time-dependent rotation~\cite{Nalbach_PRA14} around $\PauliTau^y_k$, $\hat{R}_t = \exp[i\phi_t \PauliTau^y_k/2]$, with $\phi_t = \arctan(\xi_k(t)/\Delta_k)$. 
In this way, we get $\hat{R}_t^\dagger \, \widehat{H}_{\sys}^\kup(t) \, \hat{R}_t = \epsilon_t \, \PauliTau^x_k$, 
with $\epsilon_t \equiv \hbar\Lambda_t \equiv \sqrt{\xi_k^2(t) + \Delta_k^2}$.
Notice that all the quantities introduced here depend obviously on $k$, but we dropped the corresponding index for simplicity.
In order to express the $\hat{S}_k(t)$ operator in Eq.~\eqref{S_operator} in a simpler analytic form, we make two further approximations: first, we assume that the bath correlation function $C(t)$ decays to zero in a time scale $t_B \ll t-t_0$, so that the maximum of the integral can be safely extended to infinity. 
Secondly, we assume that $\widehat{H}_{\sys}^\kup(t)$ can be regarded as constant in time intervals that are comparable
to $t_B$~\cite{Arceci_PRB17, Yamaguchi_PRE17}; this allows us to approximate the coherent evolution operator
of Eq.~\eqref{eq:Ut_evol} as $\hat{U}_{0,k}(t,t-t') \approx \exp\big\{\! - \!i \widehat{H}_{\sys}^\kup(t') \,t' / \hbar \big\}$.
Eventually, we express the time evolution differential equations in the Bloch sphere representation
$\hat{\tilde{\rho}}_{\sys}^\kup(t) = \hat{R}_t^\dagger \, \hat{\rho}_{\sys}^\kup(t) \, \hat{R}_t = \frac{1}{2} \big[ \id + \sum_{\nu}r_{\nu}(t) \PauliTau^{\nu}_k \big]$ 
with $\nu=x,y,z$, and $\id$ being the $2 \times 2$ identity matrix, so that we finally have:
\begin{equation} \label{QME_noRWA}
\left\{ \begin{array}{lcl}
	\dot{r}_x &=& -\gamma_{\Relax} (r_x - \overline{r}_x) + (\dot{\phi}_t + \gamma_{xz}) r_z \\
	\dot{r}_y &=& -\left(\gamma_{\Deco} + \displaystyle \frac{\gamma_{\Relax}}{2}\right )\, r_y - 2\Lambda_t r_z \\
	\dot{r}_z &=& -\dot{\phi}_t r_x -\gamma_{zx}(r_x - \overline{r}_x) + 2\Lambda_t r_y 
                           -\displaystyle \left(\gamma_{\Deco} - \frac{\gamma_{\Relax}}{2} \right) \, r_z 
\end{array}
\right.
\end{equation}
with $\overline{r}_x(t) = -\tanh [ \beta \epsilon_t ]$ being the ``instantaneous'' putative
equilibrium value that $r_x$ would reach in absence of the driving. 
The various (time-dependent) rate constants include the usual 
``relaxation'' $\gamma_{\Relax}$, ``pure dephasing'' $\gamma_{\Deph}$ and 
``decoherence'' $\gamma_{\Deco}$ rates~\cite{Schoen_PhyScr02}
\begin{subequations}
\begin{align}
  &\gamma_{\Relax}(t) = \frac{2\pi}{\hbar^2} \coth \left(\beta \hbar \Lambda_t \right) J(2\Lambda_t) \, \cos^2(\phi_t) \,,
  \\
  &\gamma_{\Deph}(t) = \frac{8\pi\alpha}{\hbar\beta} \, \sin^2(\phi_t) \,,
  \\
  &\gamma_{\Deco}(t) = \gamma_{\Deph}(t)+\frac{1}{2} \gamma_{\Relax}(t) \,,
\end{align}
\end{subequations}
as well as the following two extra terms
\begin{subequations}
\begin{align}
&\gamma_{zx}(t) = -\frac{\pi}{\hbar^2} \coth \left(\beta \hbar \Lambda_t \right) 
J(2\Lambda_t) \, \sin 2\phi_t \,, \\
&\gamma_{xz}(t) = \frac{4\pi\alpha}{\hbar\beta} \, \sin 2\phi_t \,.
\end{align}
\end{subequations}

If we were to neglect the unitary part of the evolution in the QME, and consider the purely dissipative QME
of Eq.~\eqref{Bloch-Redfield-diss}, 
we would have to integrate the following differential equations for the corresponding Bloch vector ${\mathbf r}^{\diss}(t)$:
\begin{equation} 
\left\{ \begin{array}{lcl}
	\dot{r}_x^{\diss} &=& -\gamma_{\Relax} (r_x^{\diss} - \overline{r}_x) + \gamma_{xz} r_z^{\diss} \\
	\dot{r}_y^{\diss} &=& -\left(\gamma_{\Deco} + \displaystyle \frac{\gamma_{\Relax}}{2}\right )\, r_y^{\diss} \\
	\dot{r}_z^{\diss} &=& -\gamma_{zx}(r_x^{\diss} - \overline{r}_x)  
                           -\displaystyle \left(\gamma_{\Deco} - \frac{\gamma_{\Relax}}{2} \right) \, r_z^{\diss} \,.
\end{array}
\right. 
\end{equation}

\section{Thermal defects density calculation} \label{appB}
%
In this appendix, we compute analytically the equilibrium thermal defects density
for an ordered transverse-field Ising chain with a fixed $h$. 
%

To start, recall that the Hamiltonian in Eq.~\eqref{H_XY} conserves the parity of the number of up (or down) spins. 
As a consequence, the Hilbert space can be partitioned into even and odd parity sectors. 
This partitioning survives also when moving to the spinless fermions picture, so that we can think to the Hamiltonian in Eq.~\eqref{eqn:HsysF} 
as the {\em even fermion} block of the total Hamiltonian $\Ham_{\rm even}^F \oplus \Ham_{\rm odd}^F$.
Observe that, considering $\Ham_{\rm even}^F$ only, we account for $N/2$ two-level systems, hence a total of $2^{N/2}$ states. 
%
Let us, for a moment, assume that we treat these $N/2$ two-level systems in a thermal state at temperature $T_b$. 
For a given momentum $k$, we reduce the basis states to just absence/presence of pairs of opposite momentum and diagonalize the 
Hamiltonian $\Ham_{\sys}^\kup=\xi_k \PauliTau^z_k + \Delta_k \PauliTau^x_k$ to get
\begin{equation}
	\Ham^\kup_{\diag} = 
	\begin{bmatrix}
		\epsilon_k && 0 \\
		0 && -\epsilon_k
	\end{bmatrix}
\end{equation}
where $\epsilon_k = \sqrt{\xi_k^2 + \Delta_k^2}$. Therefore, the corresponding thermal state is given by
\begin{align}
\begin{split}
	\hat \rho_{\thermal}^\kup &\equiv \frac{ e^{-\beta_b \Ham^\kup_{\diag}} }{ \Tr\big\{ e^{-\beta_b \Ham^\kup_{\diag}} \big\} } = \\
	&= \frac{1}{\nep^{\beta_b \epsilon_k} + \nep^{-\beta_b \epsilon_k}} 
	\begin{bmatrix}
	  \nep^{-\beta_b \epsilon_k} && 0 \\
	  0 && \nep^{\beta_b \epsilon_k}
	\end{bmatrix} .
\end{split}
\end{align}
This state is expressed in the basis of the eigenstates of $\Ham_{\sys}^\kup$, which are combinations of the original basis states
$|\!\!\uparrow_k\rangle\equiv \opcdag{k}\opcdag{-k}|0\rangle$, and $|\!\!\downarrow_k\rangle\equiv|0\rangle$. 
The corresponding creation operators $\opetadag{k}$, in terms of which 
$\Ham^\kup_{\diag}=\epsilon_k (\opetadag{k} \opeta{k}- \opeta{-k} \opetadag{-k})$, are simply related to the original fermionic operators by
\begin{equation}
\opc{k} = u_k \opeta{k} -v_k \opetadag{-k} \;,
\end{equation}
where $(u_k, v_k) = ( \epsilon_k+\xi_k,\Delta_k)/\sqrt{2\epsilon_k(\epsilon_k + \xi_k)}$.
%
Writing the defect density operator $\hat{n}_{\defects}^\kup$ in Eq.~\eqref{ndef_formula}
in terms of the $\opetadag{k}$, the corresponding expectation value over the thermal state finally reads:
\begin{align}	\label{ndef_thermal}
\begin{split}
	n_{\defects}^{T_b} &= \frac{1}{N}\sum_{k>0} \Tr\big\{ \hat{n}_{\defects}^{\kup} \, \hat \rho_{\thermal}^{\kup} \big\} \\
	&= \frac{1}{N} \sum_{k>0} \Big[ 1 - y_k \big( 1 - 2 \; \Tr\big\{ \opetadag{k} \opeta{k} \hat \rho_{\thermal}^{\kup} \big\} \big) \Big] \\
	&= \frac{1}{N} \sum_{k>0} \Big[ 1 - y_k \tanh\big( \beta_b \epsilon_k \big) \Big] \,,
\end{split}
\end{align}
where $y_k \equiv ( \Delta_k \sin k -\xi_k \cos k)/\epsilon_k$ and
$\Tr\big\{ \opetadag{k} \opeta{k} \hat \rho_{\thermal}^{\kup} \big\} = f_F(2 \beta_b \epsilon_k)$,
with $f_F(x) = 1/(1+e^x)$ being the Fermi distribution function.
Notice the factor $2$ in the Fermi function argument, due to the fact that excitations here consist of two fermions, and cost an energy $2\epsilon_k$. 
%
Eq.~\eqref{ndef_thermal} gives the density of defects for a system that thermalizes with a bath at temperature $T_b$,
but can only explore states with pairs of fermions with opposite momenta. 

The original problem, however, was a transverse-field Ising chain, and we are evidently making violence to the correct 
thermodynamics by looking only at the even-fermion sector of the Hilbert space: the counting of states, 
$2^{N/2}$, as opposed to the $2^N$ states of the full Hilbert space, is a clear witness of that error. 
Thinking in terms of the correct approach to the problem, one would immediately realize
that the very fact that the fermionic boundary conditions, 
and hence the required $k$-vectors, change when the fermionic parity changes, brings a non-trivial ``interaction''
between fermions, which does not allow for a simple thermodynamical free-fermion calculation.
However, one can devise the following shortcut, which should be correct in the thermodynamic limit $N\to \infty$, 
when the difference in the $k$-vectors associated to the two parity sectors is negligible. Let us assume that we keep 
the $N/2$ $k$-vectors fixed to those selected by the ABC boundary conditions for fermions, but allow also for the 
{\em singly occupied} states $\opcdag{k}|0\rangle$ and $\opcdag{-k}|0\rangle$.
For each of the $N/2$ values of $k$, we have $4$ states, hence $4^{N/2}=2^N$ states in total.  
%
%
The Hamiltonian at fixed $k$, in the basis given by 
$\{\opcdag{k}|0\rangle, \opcdag{-k}|0\rangle, \opcdag{k}\opcdag{-k}|0\rangle, |0\rangle \}$, 
is now four-dimensional, and given by:
\begin{equation}
	\Ham_{\full}^\kup = 
	\begin{bmatrix}
		0 && 0 && 0 && 0 \\
		0 && 0 && 0 && 0 \\
		0 && 0 && \epsilon_k && 0 \\
		0 && 0 && 0 && -\epsilon_k
	\end{bmatrix} \,.
\end{equation}
To get the thermal equilibrium state, we exponentiate $\Ham_{\full}^\kup$:
\begin{equation}
	\hat \rho_{\full}^\kup = 
	\frac{1}{Z_{\full}}
	\begin{bmatrix}
		1 && 0 && 0 && 0 \\
		0 && 1 && 0 && 0 \\
		0 && 0 && \nep^{\beta \epsilon_k} && 0 \\
		0 && 0 && 0 && \nep^{-\beta \epsilon_k}
	\end{bmatrix}
\end{equation}
where $Z_{\full} = 2 + \nep^{-\beta\epsilon_k} + \nep^{\beta \epsilon_k}$. 

Building on this result, we can compute the defect density starting from the second line in Eq.~\eqref{ndef_thermal}, 
but noting that now we have: $\Tr\big\{ \opetadag{k} \opeta{k} \hat \rho_{\full}^{\kup} \big\} = f(\beta \epsilon_k)$.
Therefore, it follows that
\begin{equation} \label{eqn:n_full}
	n_{\defects}^{\full} = \frac{1}{N} \sum_{k>0} \Big[ 1 - y_k \tanh\Big( \frac{\beta\epsilon_k}{2} \Big) \Big] \,,
\end{equation}
where $y_k$ is defined exactly as before. 
A comparison of this equation with Eq.~\eqref{ndef_thermal} shows that, restricting to states
  with only pairs of fermions with opposite momenta, the density of defects at thermal equilibrium corresponds
to the true thermodynamic one, provided the temperature of the bath is rescaled by a factor 2: $T=T_b/2$. 

\begin{figure}[b]
\centering
	\includegraphics[width=\columnwidth]{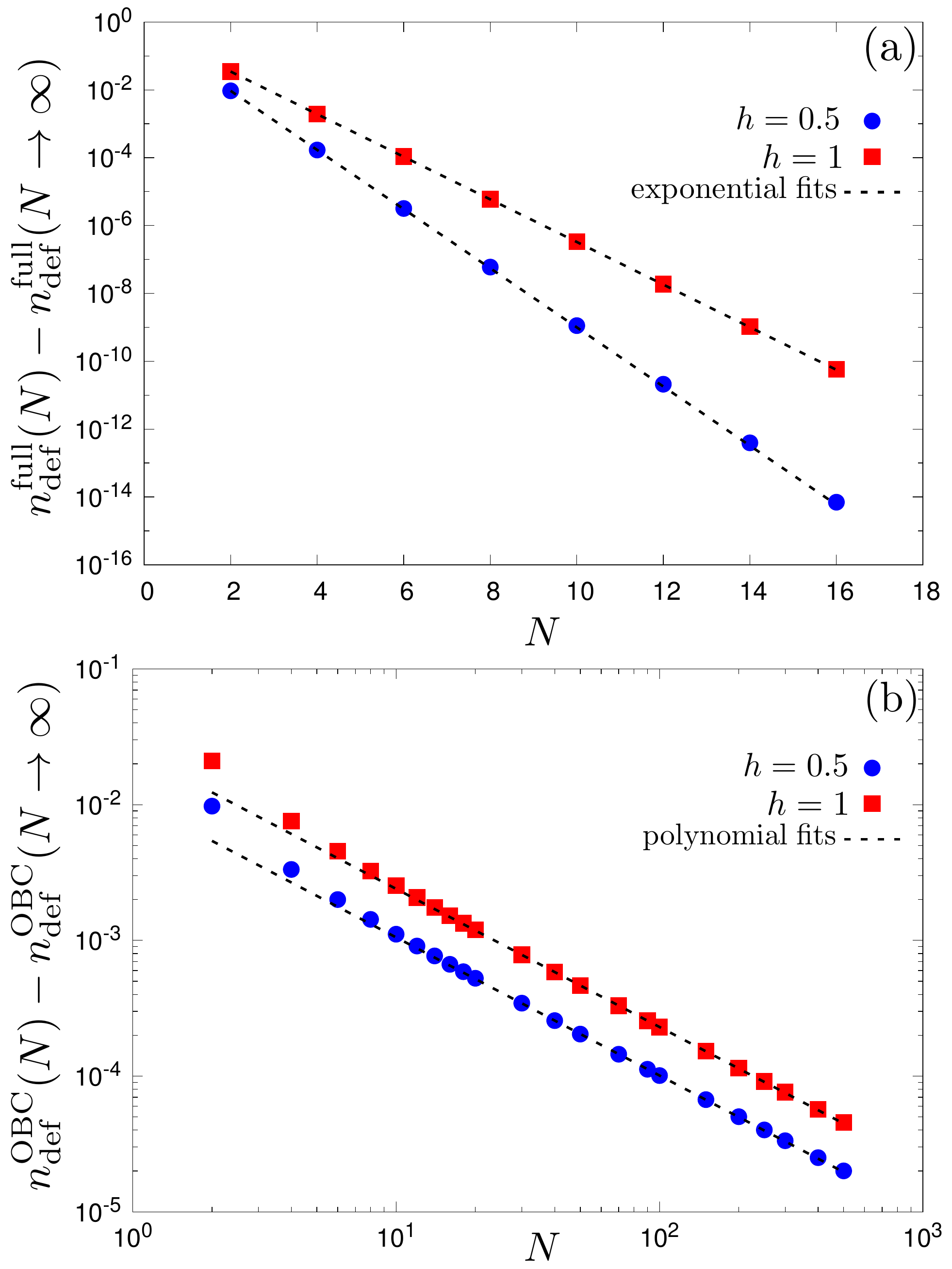}
	\caption{
	Difference between density of defects for (a) PBC and (b) OBC and the expected thermal value at the thermodynamic limit, 
	plotted versus the number of sites $N$, for $h = 0.5, 1$.
	Panel (a):  the scaling is exponential in $N$.
	Panel (b):  the scaling is polynomial in $N$; our best fit gives a convergence with $1/N$.
	}
	\label{fig:relaxation_scaling}
\end{figure}

Properly speaking, the expression in Eq.~\eqref{eqn:n_full} is exact only in the thermodynamic limit $N\to \infty$. 
One might wonder how close it describes the equilibrium thermodynamics for a finite value of $N$. 
Here, an exact and consistent reference value can be easily obtained for an Ising chain with open boundary conditions (OBC), 
where the spectrum does not depend on the fermionic parity. 
The price to be paid is that the diagonalization is not a trivial $k$-sum of $2\times 2$ problems. 
Nevertheless, for a given value of the transverse field $h$, the problem can be always reduced
to an ensemble of $N$ two-level systems.
The standard result is then~\cite{Lieb_AP61, Campostrini_JSTAT15}
\begin{equation}	\label{H_OBC}
	\Ham_{\OBC} = \sum_{m=1}^N \enOBC_m \Big[ \opetatildag_m \opetatil_m - \opetatil_m \opetatildag_m \Big]
\end{equation}
where $\opetatildag_m$ are the creation operators for the eigenstates with energies $\enOBC_m$, defined as
\begin{equation}
	\opetatil_m = \sum_{i=1}^N \big( g_{m,i} \opc{i} + h_{m,i} \opcdag{i} \big)
\end{equation}
The real coefficients $g_{m,i}, h_{m,i}$, together with the energies $\enOBC_m$,
can be computed numerically~\cite{Lieb_AP61, Campostrini_JSTAT15}.
The thermal state is thus the normalized matrix exponential of Eq.~\eqref{H_OBC}: 
\begin{equation}	\label{rhoth_OBC}
	\tilde{\rho}_{\thermal}^{(m)} = \frac{1}{\nep^{\beta \enOBC_m} + \nep^{-\beta \enOBC_m}}
	\begin{bmatrix}
		\nep^{-\beta \enOBC_m} && 0 \\
		0 && \nep^{\beta \enOBC_m}
	\end{bmatrix} \,.
\end{equation}
\begin{figure}[!t]
\centering
	\includegraphics[width=\columnwidth]{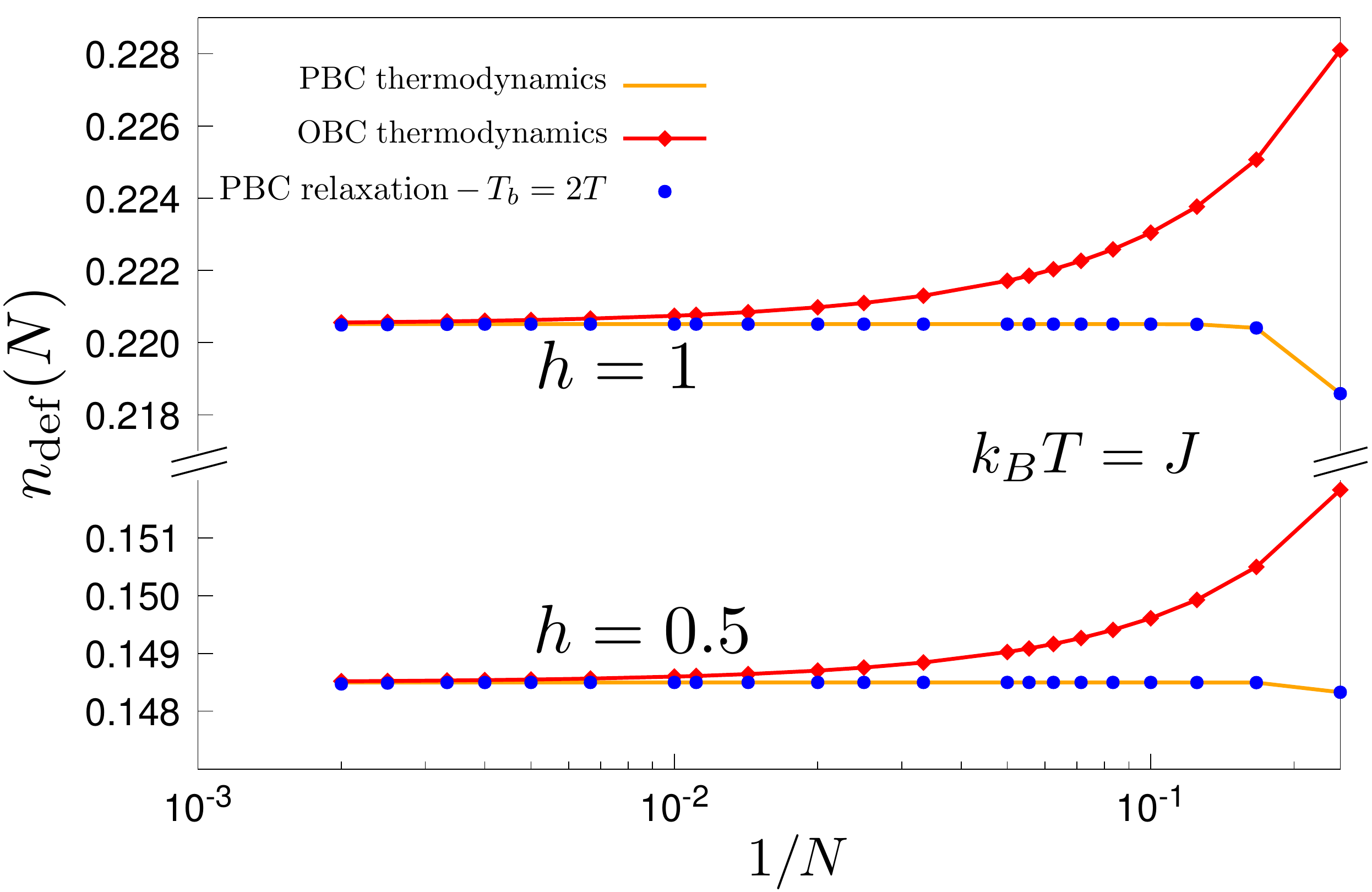}
	\caption{Thermodynamics of the defects density in the transverse-field Ising chain for $k_BT=J$. 
	The PBC thermodynamics (orange solid lines) is calculated with Eq.~\eqref{eqn:n_full}, while 
	the OBC thermodynamics (red solid lines and diamonds) corresponds to Eq.~\eqref{eqn:n_obc}. 
	The blue solid circles are PBC-QME relaxation dynamics data for $N/2$ two-level systems for $T_b=2T$.}
	\label{fig:relaxation}
\end{figure}
We can finally express the defect density operator in Eq.~\eqref{ndef_spin} by using the $\opetatil_m$ operators, 
and then compute its expectation value on the thermal state~\eqref{rhoth_OBC}: 
\begin{equation} \label{eqn:n_obc}
	n_{\defects}^{\OBC} = \frac{1}{2} - \frac{1}{N-1} \sum_{m=1}^N \Big( \tilde{A}_m \, f_F(2\beta \enOBC_m) + 
	\tilde{B}_m \, f_F(-2\beta \enOBC_m) \Big)
\end{equation}
with
\begin{subequations}
\begin{align}
	\tilde{A}_m &= \sum_{i=1}^{N-1} g_{m,i} (g_{m,i+1} + h_{m,i+1}) \,,\\
	\tilde{B}_m &= \sum_{i=1}^{N-1} h_{m,i} (g_{m,i+1} + h_{m,i+1}) \,.
\end{align}
\end{subequations}
%
%
In Fig.~\ref{fig:relaxation} we show the results for the density of defects at $k_BT=J$ versus $N$, 
for two different values of the transverse field: the critical value $h=h_c=1$, and a value in the ferromagnetically ordered phase,
$h=0.5<h_c$. The plot reports the results obtained by three different approaches:
{\em i)} the PBC formula in Eq.~\eqref{eqn:n_full} (orange solid lines); {\em ii)} an explicit QME relaxation with $T_b=2T$
(blue circles);
{\em iii)} the exact OBC evaluation, following Eq.~\eqref{eqn:n_obc} (red solid lines and diamonds). 
Notice that the convergence of the PBC results to the thermodynamic limit is exponentially fast in $N$, while the OBC data
show $1/N$ finite-size scaling corrections.
This is illustrated in Fig.~\ref{fig:relaxation_scaling}, where we show the finite-size scaling of both the PBC data (top) and 
OBC results (bottom) to the common thermodynamical limit for $k_B T = J$ and $h = 0.5, 1$.
%
%
%


\end{document}